\documentclass[final,numberedheadings,epsfig]{aipproc}
\layoutstyle{6x9}
\newcommand{\Np}{${N_{n-part}}$}
\newcommand{\Nq}{${N_{q-part}}$}
\newcommand{\be}{\begin{equation}}
\newcommand{\ee}{\end{equation}}
\newcommand{\bea}{\begin{eqnarray}}
\newcommand{\eea}{\end{eqnarray}}

\newcommand{ \dn }{${\rm dN_{ch}/d\eta}$}

\begin{document}
\vglue -1cm 
\title{Similarity of Initial States in A+A and p+p Collisions
 in Constituent Quarks Framework}
\author{Rachid NOUICER}{ address={Chemistry Department, Brookhaven
  National Laboratory, Upton, NY 11973-5000, USA} }
\vspace*{-0.3cm}\begin{abstract} The multiparticle production results
from A+A and p(${\rm \bar{p}}$)+p collisions have been compared based
on the number of nucleon participants and the number of constituent
quark (parton) participants.  In both normalizations, we observe
that the charged particle densities in Au+Au and Cu+Cu collisions are
similar for both ${\rm \sqrt{s_{NN}}}$ = 62.4 and 200 GeV.  This
implies that in symmetric nucleus-nucleus collisions the charged
particle density does not depend on the size of the two colliding
nuclei but only on the collision energy. In the nucleon participants
framework, the particle density at mid-rapidity as well as in the
limiting fragmentation region from A+A collisions are higher than
those of p(${\rm \bar{p}}$)+p collisions at the same energy. Also the
integrated total charged particle in A+A collisions as a function of
number nucleon participants is higher than p(${\rm \bar{p}}$)+p
collisions at the same energy indicating that there is no smooth
transition between peripheral A+A and nucleon-nucleon collisions.
However, when the comparison is made in the constituent quarks
framework, A+A and p(${\rm \bar{p}}$)+p collisions exhibit a striking
degree of agreement. The observations presented in this paper imply
that the number of constituent quark pairs participating in the
collision controls the particle production. One may therefore
conjecture that the initial states A+A and p+p collisions are similar
when the partonic considerations are used in normalization.  Another
interesting result is that there is an overall factorization of \dn\ shapes
as a function of collision centrality between Au+Au and Cu+Cu
collisions at the same energy, ${\rm \sqrt{s_{NN}}}$~=~200~GeV.
\end{abstract}
\maketitle
\section{Introduction}
Since the first collisions were delivered at the
Relativistic Heavy Ion Collider (RHIC), the experiments have obtained
extensive results on multiparticle production in both nucleus-nucleus
(A+A) and nucleon-nucleon (p (${\rm \bar{p}}$)+p) collisions at the same
energies~\cite{Ref1}.  However, several aspects of the comparison
between A+A and p(${\rm \bar{p}}$)+p collisions are not well
understood. For example it has been found that the particle density
per participant pair, \Np/2, in A+A collisions is substantially higher
than in p(${\rm \bar{p}}$)+p collisions at ${\rm \sqrt{s_{NN}}}$ = 200
GeV \cite{Au200}. It has also been observed that the integrated total charged
particle production, per participant nucleon pair, as a function of
\Np is essentially constant and is higher than for p(${\rm
\bar{p}}$)+p collisions at the same energy \cite{dAu} indicating that there
is no smooth transition between the two systems. These comparisons
are, however, based on scaling with the number of nucleon
participants. In the following, I will show that the A+A and p(${\rm
\bar{p}}$)+p collisions have similar initial states if the results are
scaled instead by the number of constituent quark participants,
\Nq. Within this framework, the similarity between A+A and p(${\rm \bar{p}}$)+p
collisions will be explored through global observables, which reflect
the initial state of the system.
\begin{figure}
  \includegraphics[height=.23\textheight]{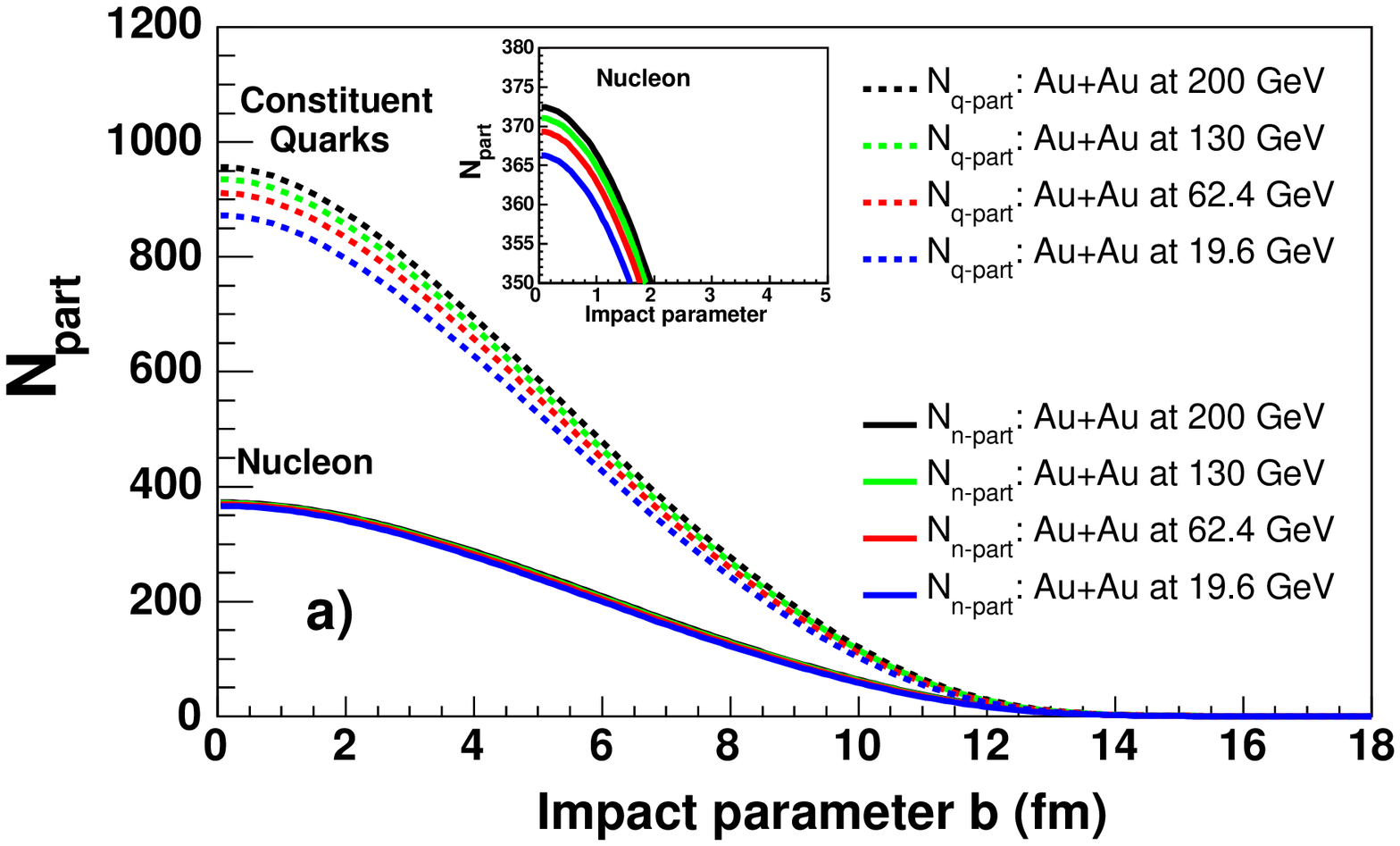}\hfill
  \includegraphics[height=.22\textheight,
  width=.3\textheight]{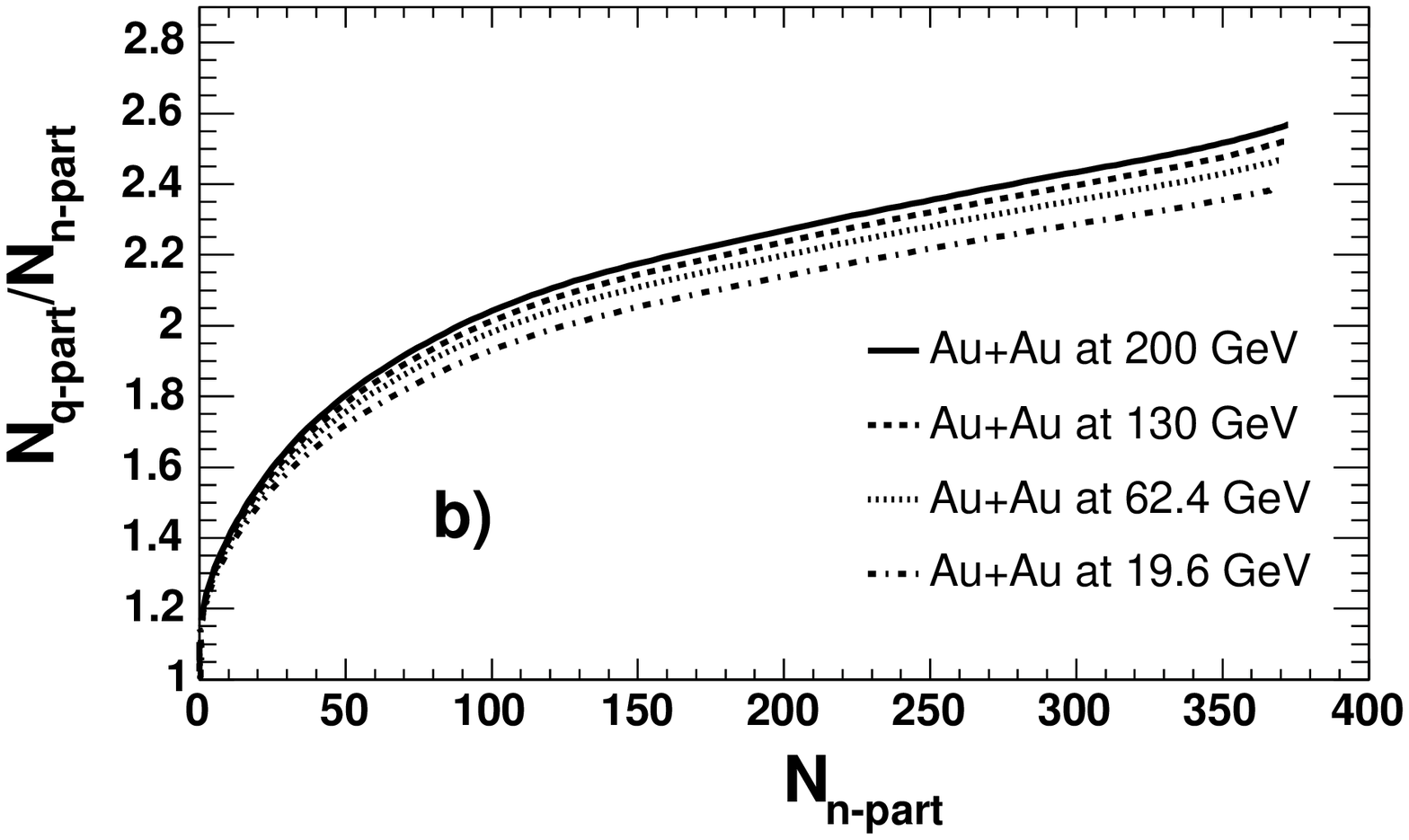}
\vspace*{-3.5cm}
  \caption{Panel a): number of nucleon participants (denoted N$_{\rm
n-part}$: solid curves) and number of constituent quark
participants (denoted N$_{\rm q-part}$: dashed curves) as a function of
the impact parameter of Au+Au collisions at ${\rm \sqrt{s_{NN}}}$ =
19.6, 62.4, 130 and 200 GeV. The inset figure represent a zoom on the
solid curves. Panel b): ratio of \Nq/\Np\
as a function of the number of nucleon participants.}
\label{fig1}
\end{figure}
\vspace*{-0.3cm} 
\section{Calculation of the number of participants}
The constituent quark (parton) model has been
introduced in Refs.~\cite{Bia}. The present work is
a continuation of the study started by Ref.~\cite{Erm} which extends to 
global observables, namely the comparison of particle density and limiting
fragmentation scaling in A+A and p(${\rm \bar{p}}$)+p collisions.

The number of nucleon participants, \Np\, and the number of constituent 
quark participants, \Nq\, are estimated using the nuclear overlap model in 
a manner similar to that used in Ref.~\cite{Erm}. The nuclear density 
profile is thus assumed to have a  Woods-Saxon form, 
\vspace*{-0.3cm}\begin{equation}
	n_{A}(r) =   \frac{n_{0}}{1+exp[(r-R)/d]},   
\end{equation} 
where $n_{0}$ = 0.17 $fm^{-3}$, $R=(1.12~A^{1/3} - 0.86~A^{-1/3}$) fm and 
$d$ = 0.53 fm.

The number of nucleon participants, ${N_{n-part}}$, for
nucleus-nucleus (A+B) collisions is calculated using the relation,
\vspace*{-0.4cm}\bea \normalsize N_{n-part}|_{AB} = \int
d^{2}sT_{A}(\vec{s}) \{ 1 -
\bigg[1-\frac{\sigma^{inel}_{NN}T_{B}(\vec{s}-\vec{b})}{B}
\bigg]^{B}\} \\ \nonumber + \int d^{2}sT_{B}(\vec{s}-\vec{b}) \{1 -
\bigg[1-\frac{\sigma^{inel}_{NN}T_{A}(\vec{s})}{A}\bigg]^{A} \} \eea
where {\small $T(b)~=~\int_{- \infty}^{+ \infty} dz n_{A}(
\sqrt{b^{2}+z^{2}})$}, is the thickness function. $A$ and $B$ are the
mass number of the two colliding nuclei and the ${\sigma^{inel}_{NN}}$
is the inelastic nucleon-nucleon cross section.  

The number of constituent quark participants, \Nq, is calculated in a
similar manner by taking into account the following changes related to
the physical realities: 1) the density is three times that of nucleon
density with $n_{0}^{q}$ = 3$n_{0}$ = 0.51 $fm^{-3}$; 2) the cross
sections $\sigma_{qq}$ = $\sigma^{inel}_{NN}$/9; 3) the mass numbers
of the colliding nuclei are three times their values, keeping the size
of the nuclei same as in the case of \Np. For p(${\rm \bar{p}}$)+p
collisions the same procedure has been used to calculate the number of
constituent quark participants by using $A$ = 3 and $B$ = 3 and
considering nucleons as hard spheres of uniform radii of 0.8
fm~\cite{Wong}.
\begin{table}[]
\begin{tabular}{lrrrrrrrrrr}
\hline
 ${\rm \sqrt{s_{NN}}}$  & 19.6& 53 & 56 & 62.4& 130 & 200 & 540 & 630 & 900 & 1800\\
\hline
$\sigma^{inel}_{NN}$ &31.5 & 35.0& 35.3 &36.0 &39.3  &42.0 &48.0  &48.6&51.0 &56.0 \\
\hline
\end{tabular}
\caption{Inelastic cross section for nucleon-nucleon collisions
($\sigma^{inel}_{NN}$) as function of colliding energy. My calculation
for $\sigma^{inel}_{NN}$ adopt similar manner as in Ref.~\cite{Bha}}.
\label{tab1}
\end{table}
\begin{figure}[]
\includegraphics[height=.2\textheight]{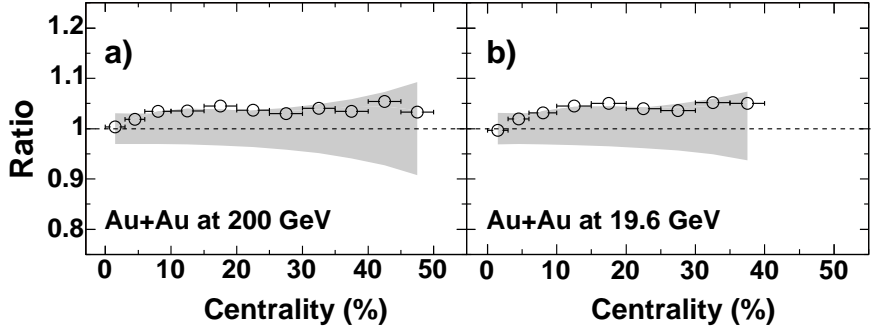}
  \caption{Quantitative evaluation of the model calculations expressed
as the ratio of the average number of nucleon participants of the
PHOBOS Glauber calculations~\cite{Bac1} to the present work as a function of
centrality. Panel a) and b) correspond to Au+Au
collisions at ${\rm \sqrt{s_{NN}}}$ = 200 and 19.6 GeV, respectively.
The gray bands correspond to the
systematic errors on the ${\rm \langle N_{n-part} \rangle}$ of PHOBOS
Glauber calculations.}
\vspace*{-4cm}
\label{fig2}
\end{figure}

Figure~\ref{fig1}a) shows the number of nucleon (solid curves) and
constituent quark (dashed curves) participants for Au+Au collisions
as a function of the impact parameter. Figure~\ref{fig1}b) presents
the ratio of \Nq/\Np\ as a function of \Np\ for Au+Au collisions at
RHIC energies. The ratio shows that the correlation between \Np\ and
\Nq\ is not linear and that it depends slightly on the colliding
energy. The inelastic nucleon-nucleon cross sections,
$\sigma^{inel}_{NN}$, used in the present work are listed in
TABLE~\ref{tab1}.

Figure~\ref{fig2} presents the ratio of the \Np\ from the PHOBOS
Glauber calculation based on HIJING~\cite{Bac1} to the present
calculation and shows good agreement (within systematic
errors). 

\begin{figure}[]
\caption{ Particle density per constituent quark pair (open symbols)
and particle density per nucleon participant pair (solid points)
produced in central (6\%) nucleus-nucleus (A+A) collisions as function
of collision energy at AGS, SPS~\cite{Ahl,Bach,Blu} and
RHIC~\cite{Gunt} and in ${\rm p(\bar{p})+p}$ collisions at ISR
energies~\cite{Abe} and ${\rm {p}+p}$ at 200 GeV at
RHIC~\cite{Brah1}. The errors bars correspond to the systematic
errors. The solid line represents a linear fit through solid points
for A+A data, ${\rm f_{AA} = -0.287 + 0.757 ln (\sqrt{s}) }$. The
dashed~dotted~line corresponds to the fit through solid points of
${\rm p(\bar{p})+p}$ collisions, ${\rm f_{pp} = 2.25-0.41
ln(\sqrt{s})+ 0.09 ln^{2}(\sqrt{s})}$.  The dashed line corresponds to
a linear fit trough the open symbols in the constituent quarks
framework, ${\rm f_{ p(\bar{p})p/AA} = -0.02+0.27ln(\sqrt{s})}$.  }
\includegraphics[width=6cm]{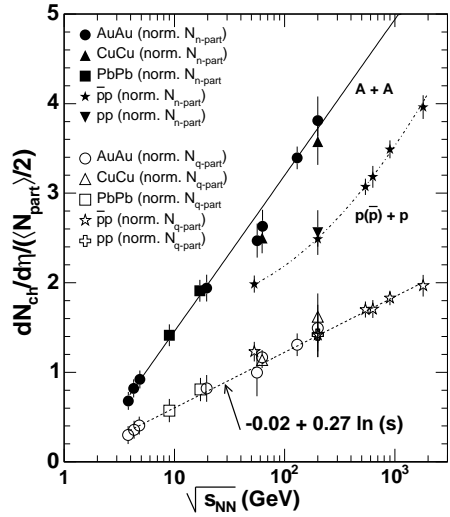}
\label{fig3}
\end{figure}

\begin{figure}[t]
  \includegraphics[height=.35\textheight]{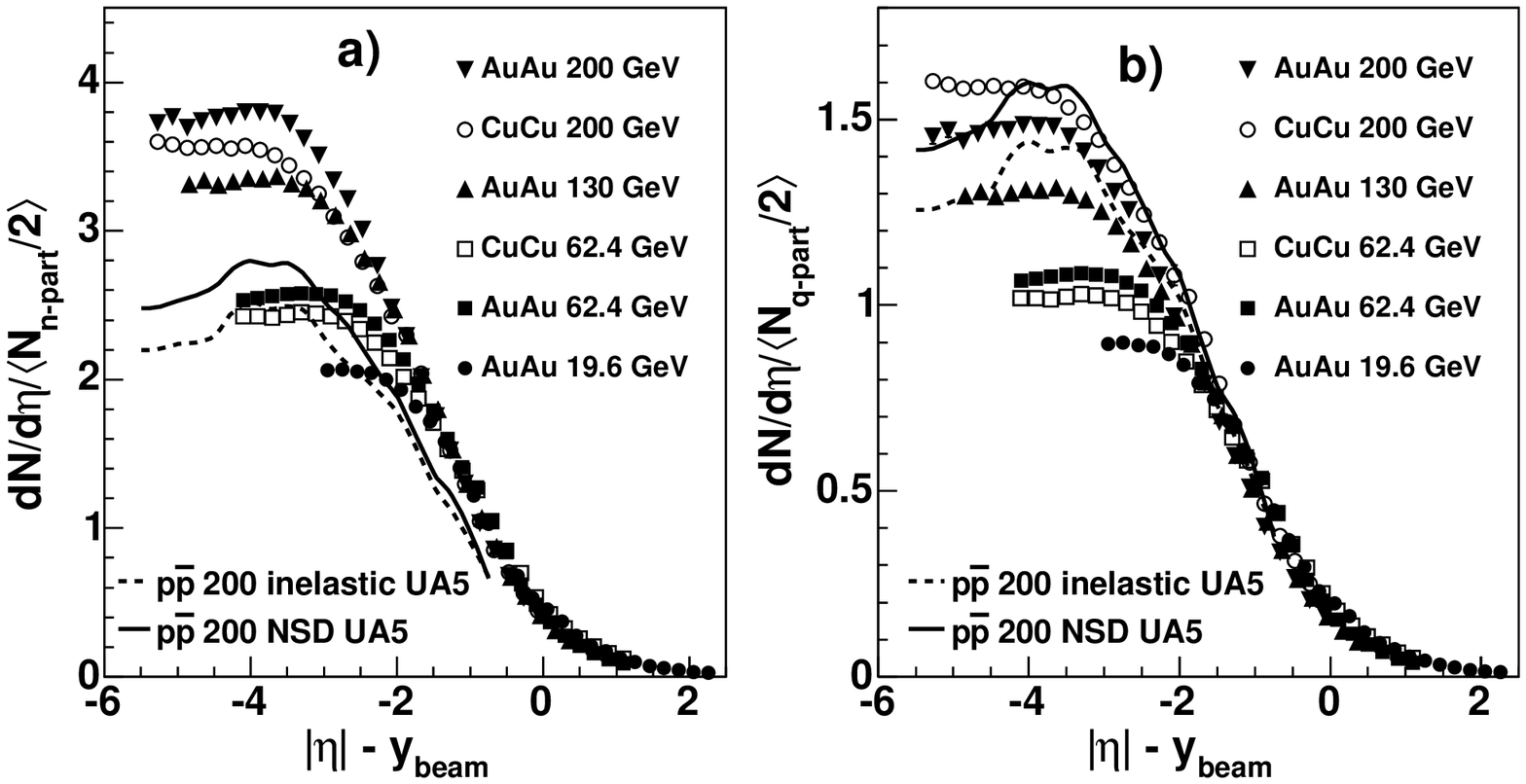}
  \caption{Pseudorapidity distributions of charged particle for Au+Au,
   Cu+Cu~\cite{Gunt} collisions at RHIC energies compared to p(${\rm
   \bar{p}}$)+p collisions at 200 GeV~\cite{Abe}.  The distributions have been
   shifted to ${\rm \eta -y_{beam}}$ in order to study the
   fragmentation regions in one of the nucleus rest frame.  Panels a)
   and b) correspond to \dn\ distributions scaled to the number of
   nucleon participants and to the number of constituent quark
   participants, respectively.  For clarity the systematic errors have
   been removed.
} 
  \label{fig:4}
\end{figure}
\vspace*{-0.3cm}
\section{Charged Particle Densities}
Figure~\ref{fig3} shows the primary charged particle density for
central collisions at mid-rapidity divided by the number of
participant nucleon pairs (\Np/2) and participant constituent quark
pairs (\Nq/2) as solid and open symbols, respectively. The plotted data
are for Au+Au collisions at AGS, Pb+Pb collisions at the CERN
SPS~\cite{Ahl,Bach,Blu} and for Au+Au and Cu+Cu collisions at
RHIC~\cite{Gunt}. Also shown for comparison are results from p(${\rm
\bar{p}}$)+p collisions data \cite{Abe, Brah1}.

The particle density per nucleon participant pair for A+A collisions
(solid points) shows an approximately logarithmic rise with
$\sqrt{s_{_{NN}}}$ over the full range of collision energies.  The
comparison of the particle density per nucleon of Au+Au to Cu+Cu
collisions at the same energies, ${\rm \sqrt{s_{NN}}}$ = 62.4 and 200
GeV indicates that in the symmetric nucleus-nucleus collisions the
density per nucleon participant does not depend on the size of the two
colliding nuclei but only on the collision energy.  This means for
Si+Si collisions at ${\rm \sqrt{s_{NN}}}$ = 200 GeV, the particle
density per nucleon participant will be similar to Au+Au collisions at
the same energy. Figure 3 shows (solid points) that the charged
particle multiplicity per participant nucleon pair in A+A collisions
is higher compared to p(${\rm \bar{p}}$)+p collisions
at the same energy.

In contrast, we observe that the multiplicity per constituent quark
participant pair (open symbols in figure~\ref{fig3}) is similar for
nucleus-nucleus collisions and nucleon-nucleon collisions at the same
energy. It thus appears that using partonic participants accounts for
the observed multiplicity in both A+A and p(${\rm \bar{p}}$)+p
collisions. One may therefore conjecture that the initial states
in A+A and p(${\rm \bar{p}}$)+p collisions are similar. 

\vspace*{-0.3cm}
\section{Extended Longitudinal Scaling}
In general, the charged particle production in the limiting
fragmentation region is thought to be distinct from that at
mid-rapidity, although there is no obvious evidence for two separate
regions at any of the RHIC energies. This observation is made based
on the \dn\ distributions of charged particle presented in Ref.~\cite{Mark1}.

Figure~\ref{fig:4} shows a comparison of the most central (6\%) Au+Au
 and Cu+Cu collisions at several RHIC energies compared to p(${\rm
 \bar{p}}$)+p ( (inelastic and non single diffractive (NSD))
 collisions at 200 GeV.  When normalized to \Np/2, (figure~4a), we
 observe that the multiplicity in the limiting fragmentation region in
 A+A collisions is higher than for p(${\rm \bar{p}}$)+p collisions.
 If, however, the comparison is carried out for multiplicities
 normalized to \Nq/2 (figure~4~b), A+A and p(${\rm \bar{p}}$)+p
 collisions exhibit a striking degree of agreement. Again, this
 observation implies that the number of constituent quark pairs
 participating in the collision controls the particle production.

\section{System size independence of pseudorapidity shapes}
\begin{figure}[t]
  \includegraphics[height=.25\textheight]{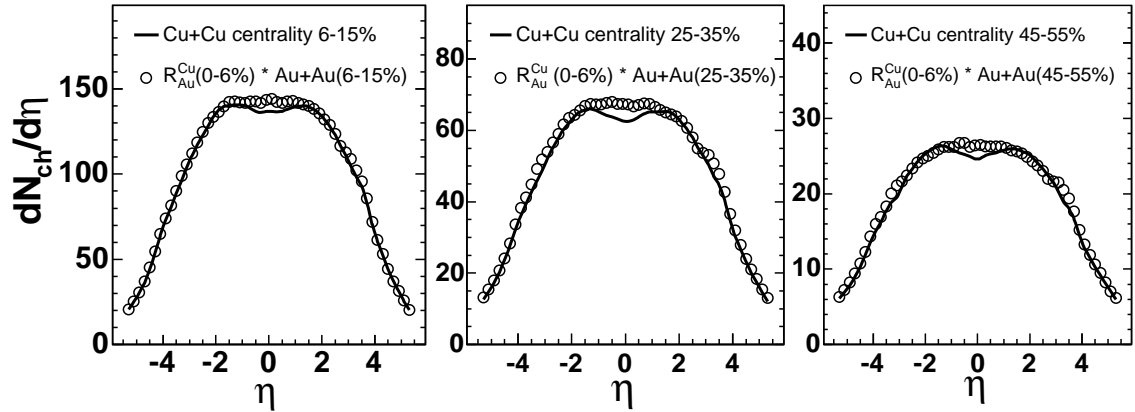}
  \caption{Comparison of \dn\ distributions of Cu+Cu to Au+Au
collisions~\cite{Gunt} at the same collision centrality and energy,
200 GeV, presented for different centrality bins. The \dn\ distributions
of Au+Au collisions have been multiplied by factor which corresponds to
the ratio of the measured \dn\ distributions of Cu+Cu central (6\%)
collisions to the measured \dn\ of Au+Au central (6\%) collisions.
For clarity, the systematic errors are not shown.}
\label{fig5}
\end{figure}
For 0-6\% central collisions 
we find that the multiplicity shapes are essentially identical for Au+Au 
and Cu+Cu collisions and differ only by a constant factor,
\begin{equation}
R^{Cu}_{Au} (0-6\%) = \frac{dN/d\eta (Cu+Cu: 0-6\%)}{dN/d\eta (Au+Au: 0-6\%)}.
\end{equation}
Figure~\ref{fig5} illustrates the fact that the same ratio between
Au+Au and Cu+Cu collisions pseudorapidity distributions holds for all
centrality bins (see figure~\ref{fig5}) such that the shapes of the
\dn\ distributions for the same centrality bin are similar for the two
systems. The small difference at mid-rapidiy can be related to the
difference of the mean $P_{T}$ of charged particle in Cu+Cu and Au+Au
collisions but it falls well within the systematic errors so this
difference is not significant. It thus appears that the \dn\
shapes are independent of the overall size of the colliding nuclei, at
least between the Cu+Cu and Au+Au systems studied here.
\vglue -3.5cm
\section{Summary}
\vspace*{-0.2cm}
I have shown that the charged particle multiplicity, both at
mid-rapidity and in the limiting fragmentation region scale with the
number of constituent quark participants, both in nucleus-nucleus
systems of different sizes and in nucleon-nucleon collisions.  This
observation implies that both the overall the particle production and
the distribution in pseudorapidity is controlled by at the participant
quark level. In addition, I have shown that shapes of the charged
particle multiplicity in Au+Au and Cu+Cu collisions at the same
energies are very similar and that they differ only by a overall
factor, even at different centralities given by the fraction of the
overall cross section.

\vspace*{-0.5cm}
\begin{theacknowledgments}
{
\vspace*{-0.4cm} I thank B. B. Back and M. D. Baker for discussions
and a careful reading of the manuscript. This work was supported by
U.S. DOE Grant No. DE-AC02-98CH10886.}
\end{theacknowledgments}
\vspace*{-0.5cm}
\bibliographystyle{aipprocl} 
\bibliography{sample}
\IfFileExists{\jobname.bbl}{}
 {\typeout{}
  \typeout{******************************************}
  \typeout{** Please run "bibtex \jobname" to optain}
  \typeout{** the bibliography and then re-run LaTeX}
  \typeout{** twice to fix the references!}
  \typeout{******************************************}
  \typeout{}
 }

\end{document}